\documentclass[iop,apjl]{emulateapj}

\usepackage{apjfonts}
\newcommand{\as}{\ensuremath{^{\prime\prime}}}                    % Arcsecond symbol
\newcommand{\am}{\ensuremath{^\prime}}                            % Arcminute symbol
\newcommand{\Teff}{\ensuremath{T_{\rm eff}}}                      % Effective temperature symbol
\newcommand{\logg}{\ensuremath{\log g}}                           % log(g) symbol
\newcommand{\FeH}{\ensuremath{\left[\frac{\rm Fe}{\rm H}\right]}} % [Fe/H] symbol
\newcommand{\MoH}{\ensuremath{\left[\frac{\rm M}{\rm H}\right]}}  % [M/H] symbol
\newcommand{\Msun}{\ensuremath{\,{\rm M}_\odot}}                  % Solar mass symbol
\newcommand{\Rsun}{\ensuremath{\,{\rm R}_\odot}}                  % Solar radius symbol
\newcommand{\psun}{\ensuremath{\,\rho_\odot}}                     % Solar density symbol
\newcommand{\Mjup}{\ensuremath{\,{\rm M}_{\rm Jup}}}              % Jupiter mass symbol
\newcommand{\Rjup}{\ensuremath{\,{\rm R}_{\rm Jup}}}              % Jupiter radius symbol
\newcommand{\pjup}{\ensuremath{\,\rho_{\rm Jup}}}                 % Jupiter density symbol
\newcommand{\mss}{\,m\,s$^{-2}$}                                  % m/s^2 symbol
\newcommand{\mc}[1]{\multicolumn{2}{c}{#1}}                       % shorthand for two-column centred
\newcommand{\reff}[1]{{#1}}                                   % makes corrections bold-face if wanted.

\newcommand{\erm}[3]{\mc{\ensuremath{#1^{+#2}_{-#3}}}}

\begin{document} %%%%%%%%%%%%%%%%%%%%%%%%%%%%%%%%%%%%%%%%%%%%%%%%%%%%%%%%%%%%%%%%%%%%%%%%%%%%%%%%%%
%%%%%%%%%%%%%%%%%%%%%%%%%%%%%%%%%%%%%%%%%%%%%%%%%%%%%%%%%%%%%%%%%%%%%%%%%%%%%%%%%%%%%%%%%%%%%%%%%%%

\title[Physical properties of WASP-18]{Physical properties of the 0.94-day period transiting planetary system WASP-18\footnote{Based on data collected by MiNDSTEp with the Danish 1.54\,m telescope at the ESO La Silla Observatory}}

\author{John Southworth\,$^1$,                                                % Keele
        T C Hinse\,$^{2,3}$,                                                  % Armagh + NBI
        M Dominik\,$^4$\footnote{Royal Society University Research Fellow},   % St.A
        M Glitrup\,$^5$,                                                      % Aarhus
        U G J{\o}rgensen\,$^3$,                                               % NBI
        C Liebig\,$^6$,                                                       % Heidelberg
        M Mathiasen\,$^4$,                                                    % St.A
        D R Anderson\,$^1$,                                                   % Keele
        V Bozza\,$^{7,8}$,                                                    % Salerno + INFN
        P Browne\,$^4$,                                                       % St.A
        M Burgdorf\,$^9$,                                                     % Stuttgart
        S Calchi Novati\,$^{7,8}$,                                            % Salerno + INFN
        S Dreizler\,$^{10}$,                                                  % Goettingen
        F Finet\,$^{11}$,                                                     % Liege
%         F Grundahl\,$^5$,                                                     % Aarhus
        K Harps{\o}e\,$^3$,                                                   % NBI
        F Hessman\,$^{10}$,                                                   % Goettingen
        M Hundertmark\,$^{10}$,                                               % Goettingen
        G Maier\,$^6$,                                                        % Heidelberg
        L Mancini\,$^{7,8}$,                                                  % Salerno + INFN
        P F L Maxted\,$^1$,                                                   % Keele
        S Rahvar\,$^{12}$,                                                    % Sharif
        D Ricci\,$^{11}$,                                                     % Liege
        G Scarpetta\,$^{7,8}$,                                                % Salerno + INFN
        J Skottfelt\,$^3$,                                                    % NBI
        C Snodgrass\,$^{13}$,                                                 % ESO
        J Surdej\,$^{11}$,                                                    % Liege
%         J Wambsganss\,$^6$,                                                   % Heidelberg
        F Zimmer\,$^6$,                                                       % Heidelberg
       }

\address{$^1$\,Astrophysics Group, Keele University, Newcastle-under-Lyme, ST5 5BG, UK \\
         $^2$\,Armagh Observatory, College Hill, Armagh, BT61 9DG, Northern Ireland, UK \\
         $^3$\,Niels Bohr Institute and Centre for Star and Planet Formation,
                                    University of Copenhagen, Juliane Maries vej 30, 2100 Copenhagen \O, Denmark \\
         $^4$\,SUPA, University of St Andrews, School of Physics \& Astronomy, North Haugh, St Andrews, KY16 9SS, UK \\
         $^5$\,Department of Physics \& Astronomy, Aarhus University, Ny Munkegade, 8000 Aarhus C, Denmark \\
         $^6$\,Astronomisches Rechen-Institut, Zentrum f\"ur Astronomie, Universit\"at Heidelberg,
                                    M\"onchhofstra{\ss}e 12-14, 69120 Heidelberg, Germany \\
         $^7$\,Dipartimento di Fisica ``E. R. Caianiello'', Universit\`a di Salerno, Baronissi, Italy \\
         $^8$\,Instituto Nazionale di Fisica Nucleare, Sezione di Napoli, Italy \\
%          $^9$\,Deutsches SOFIA Institut, Universitaet Stuttgart, Pfaffenwaldring 31, 70569 Stuttgart, Germany \\
         $^9$\,Deutsches SOFIA Institut, NASA Ames Research Center, Mail Stop 211-3, Moffett Field, CA 94035, USA \\
         $^{10}$\,Institut f\"ur Astrophysik, Georg-August-Universit\"at G\"ottingen,
                                    Friedrich-Hund-Platz 1, 37077 G\"ottingen, Germany \\
         $^{11}$\,Institut d'Astrophysique et de G\'eophysique, Universit\'e de Li\`ege, 4000 Li\`ege, Belgium \\
         $^{12}$\,Department of Physics, Sharif University of Technology, Tehran, Iran \\
         $^{13}$\,European Southern Observatory, Casilla 19001, Santiago 19, Chile
        }

% \ead{jkt@astro.keele.ac.uk}

\begin{abstract}
We present high-precision photometry of five consecutive transits of WASP-18, an extrasolar planetary system with one of the shortest orbital periods known. Through the use of telescope defocussing we achieve a photometric precision of 0.47--0.83 mmag per observation over complete transit events. The data are analysed using the {\sc jktebop} code and three different sets of stellar evolutionary models. We find the mass and radius of the planet to be $M_{\rm b} = 10.43 \pm 0.30 \pm 0.24$\Mjup\ and $R_{\rm b} = 1.165 \pm 0.055 \pm 0.014$\Rjup\ (statistical and systematic errors) respectively. The systematic errors in the orbital separation and the stellar and planetary masses, arising from the use of theoretical predictions, are of a similar size to the statistical errors and set a limit on our understanding of the WASP-18 system. We point out that seven of the nine known massive transiting planets ($M_{\rm b} > 3$\Mjup) have eccentric orbits, whereas significant orbital eccentricity has been detected for only four of the 46 less massive planets. This may indicate that there are two different populations of transiting planets, but could also be explained by observational biases. Further radial velocity observations of low-mass planets will make it possible to choose between these two scenarios.
\end{abstract}

% \begin{keywords}
% stars: planetary systems --- stars: individual: WASP-18 --- stars: binaries: eclipsing
% \end{keywords}

\maketitle

%%%%%%%%%%%%%%%%%%%%%%%%%%%%%%%%%%%%%%%%%%%%%%%%%%%%%%%%%%%%%%%%%%%%%%%%%%%%%%%%%%%%%%%%%%%%%%%%%%%
%%%%%%%%%%%%%%%%%%%%%%%%%%%%%%%%%%%%%%%%%%%%%%%%%%%%%%%%%%%%%%%%%%%%%%%%%%%%%%%%%%%%%%%%%%%%%%%%%%%
%%%%%%%%%%%%%%%%%%%%%%%%%%%%%%%%%%%%%%%%%%%%%%%%%%%%%%%%%%%%%%%%%%%%%%%%%%%%%%%%%%%%%%%%%%%%%%%%%%%

\section{Introduction}

The recent discovery of the transiting extrasolar planetary system WASP-18 \citep[][hereafter H09]{Hellier+09natur} lights the way towards understanding the tidal interactions between giant planets and their parent stars. WASP-18\,b is one of the shortest-period ($P_{\rm orb} = 0.94$\,d) and most massive ($M_{\rm b} = 10$\Mjup) extrasolar planets known. These properties make it an unparallelled indicator of the tidal dissipation parameters \citep{GoldreichSoter66icar} for the star and the planet \citep{Jackson+09apj}. A value similar to that observed for Solar system bodies ($Q \sim 10^4$--$10^6$; \citealt{Peale99araa}) would cause the orbital period of WASP-18 to decrease at a sufficient rate for the effect to be observable within ten years (H09).

In this work we present high-precision follow-up photometry of WASP-18, obtained using telescope-defocussing techniques \citep{Me+09mn,Me+09mn2} which give a scatter of only 0.47 to 0.83 mmag per observation. These are analysed to yield improved physical properties of the WASP-18 system, with careful attention paid to statistical and systematic errors. The quality of the light curve is a critical factor in measurement of the physical properties of transiting planets \citep{Me09mn}. In the case of WASP-18 the systematic errors arising from the use of theoretical stellar models are also important, and are a limiting factor in the understanding of this system.

%%%%%%%%%%%%%%%%%%%%%%%%%%%%%%%%%%%%%%%%%%%%%%%%%%%%%%%%%%%%%%%%%%%%%%%%%%%%%%%%%%%%%%%%%%%%%%%%%%%

\section{Observations and data reduction}

\begin{table*} \centering
\caption{\label{tab:obslog} Log of the observations presented in this work.
$N_{\rm obs}$ is the number of observations. `Moon' and `Distance' are the fractional
illumination of the Moon at its distance from WASP-18 at the midpoint of the transit.}
\begin{tabular}{lccccclccc} \hline \hline
Date & Start time (UT) & End time (UT) & $N_{\rm obs}$ & Exposure time (s)
                                & Filter & Airmass & Moon & Distance ($^\circ$) & Scatter (mmag) \\
\hline
2009 09 07 & 06:41 & 10:05 &  93 & 80.0 & $V$ & 1.05 $\to$ 1.27            & 0.860 & 61.9 & 0.83 \\
2009 09 08 & 05:00 & 10:04 & 140 & 80.0 & $V$ & 1.15 $\to$ 1.04 $\to$ 1.28 & 0.784 & 67.1 & 0.68 \\
2009 09 09 & 03:55 & 10:00 & 169 & 80.0 & $V$ & 1.31 $\to$ 1.04 $\to$ 1.27 & 0.696 & 73.3 & 0.51 \\
2009 09 10 & 03:25 & 08:29 & 141 & 80.0 & $V$ & 1.41 $\to$ 1.04 $\to$ 1.10 & 0.597 & 80.2 & 0.56 \\
2009 09 11 & 04:07 & 07:12 &  86 & 80.0 & $V$ & 1.25 $\to$ 1.04            & 0.492 & 87.6 & 0.47 \\
\hline \hline\\ \end{tabular} \end{table*}

We observed five consecutive transits of WASP-18 on the nights of 2009 September 7--11, using the 1.54\,m Danish Telescope at ESO La Silla with the focal-reducing imager DFOSC. The plate scale of this setup is 0.39\as\,pixel$^{-1}$. The CCD was windowed down to an area of 12\am$\times$8\am\ in order to decrease the readout time to 51\,s. This window was chosen to include WASP-18 (HD\,10069, spectral type F6\,V, $V = 9.30$, $B-V = 0.44$) and a good comparison star (HD\,10179, spectral type F3\,V, $V = 9.65$, $B-V = 0.45$).

A Johnson $V$ filter and exposure times of 80\,s were used (instead of our usual Cousins $R$ and 120\,s), to obtain lower count rates from the target and comparison stars. We defocussed the telescope to a point spread function (PSF) diameter of $\sim$90 pixels (35\as), to limit the peak counts to 45\,000 per pixel for WASP-18 and also lower the flat-fielding noise. This focus setting was used for all observations and the pointing of the telescope was maintained using autoguiding. An observing log is given in Table\,\ref{tab:obslog}.

We took several images with the telescope properly focussed in order to check that there were no nearby stars contaminating the PSF of WASP-18. There are five detectable objects nearby, but all are at least 100 pixels distant. The brightest is 6.75\,mag fainter than WASP-18 and the other four are more than 9.5\,mag fainter. We conclude that the PSF of WASP-18 is not contaminated by any star detectable with our equipment.

Data reduction was performed in the same way as in \citet{Me+09mn,Me+09mn2}. In short, we run a reduction pipeline written in {\sc idl}%
%-------------------------------------------footnote---------------------------------------------
\footnote{The acronym {\sc idl} stands for Interactive Data Language and is a trademark of ITT Visual
Information Solutions. For further details see {\tt http://www.ittvis.com/ProductServices/IDL.aspx}.}
%-------------------------------------------footnote---------------------------------------------
which uses an implementation of the {\sc daophot} package \citep{Stetson87pasp} to perform aperture photometry. For each night's data the apertures were placed interactively and their positions fixed for each CCD image. We found that the most precise photometry was obtained using an object aperture of radius 60 pixels and a sky annulus of radius 85--110 pixels. The choice of aperture sizes (within reason) excludes flux from the nearby faint stars, and has a negligible effect on the resulting photometry.

\begin{figure} \includegraphics[width=0.48\textwidth,angle=0]{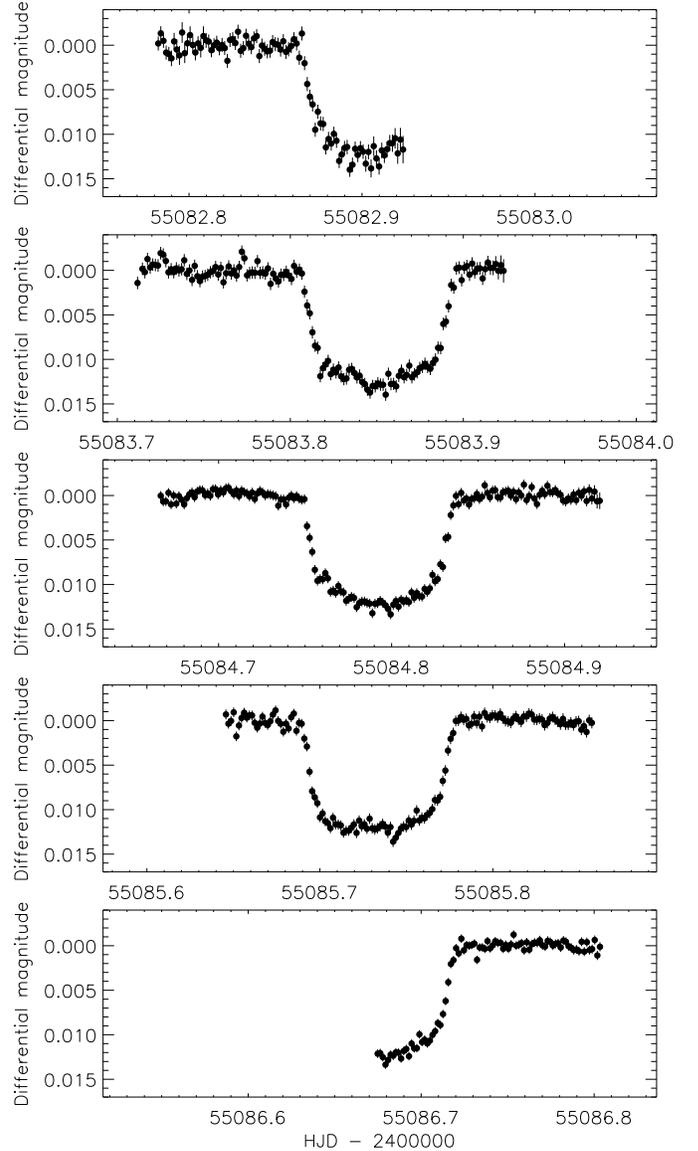}
\caption{\label{fig:plotlc} Light curves of WASP-18 from the five nights of
observations. For each night the error bars have been scaled to give
a reduced $\chi^2$ of $\chi^2_{\ \nu} = 1.0$.\\} \end{figure}

We calculated differential magnitudes for WASP-18 using an ensemble of four comparison stars, of which HD\,10179 is by far the brightest. The light curve for each night was normalised to zero differential magnitude by fitting a straight line to the observations taken outside transit, whilst simultaneously optimising the weights of the four comparison stars. We have applied bias and flat-field corrections to the images, but find that this does not have a significant effect on the photometry. The individual light curves are shown in Fig.\,\ref{fig:plotlc}, and the full 629 datapoints are given in Table\,\ref{tab:data}. The scatter in the final light curves varies from 0.47 to 0.83 mmag per point, and is higher for data taken when the moon was bright (Table\,\ref{tab:obslog}).

\begin{table} \centering
\caption{\label{tab:data} Photometric observations of WASP-18.
This table is included to show the form of the data; the complete
dataset can be found in the electronic version of this work.}
\begin{tabular}{lrr} \hline \hline
Midpoint of & Relative $R$ & Error in $R$ \\
observation (BJD) & magnitude & magnitude \\
\hline
55082.783033 & $-$0.000200 & 0.000788 \\
55082.784491 & $-$0.001340 & 0.000790 \\
55082.786030 & $-$0.000495 & 0.000786 \\
55082.787570 &    0.000825 & 0.000786 \\
55082.789109 &    0.000989 & 0.000804 \\
55082.790649 &    0.001469 & 0.000880 \\
55082.792188 & $-$0.000431 & 0.000965 \\
55082.793727 &    0.000483 & 0.001159 \\
55082.795267 &    0.001142 & 0.001067 \\
55082.796806 & $-$0.001431 & 0.001156 \\
\hline \hline\\ \end{tabular} \end{table}

%%%%%%%%%%%%%%%%%%%%%%%%%%%%%%%%%%%%%%%%%%%%%%%%%%%%%%%%%%%%%%%%%%%%%%%%%%%%%%%%%%%%%%%%%%%%%%%%%%%

\section{Light curve analysis}

\begin{figure*} \centering \includegraphics[width=1.00\textwidth,angle=0]{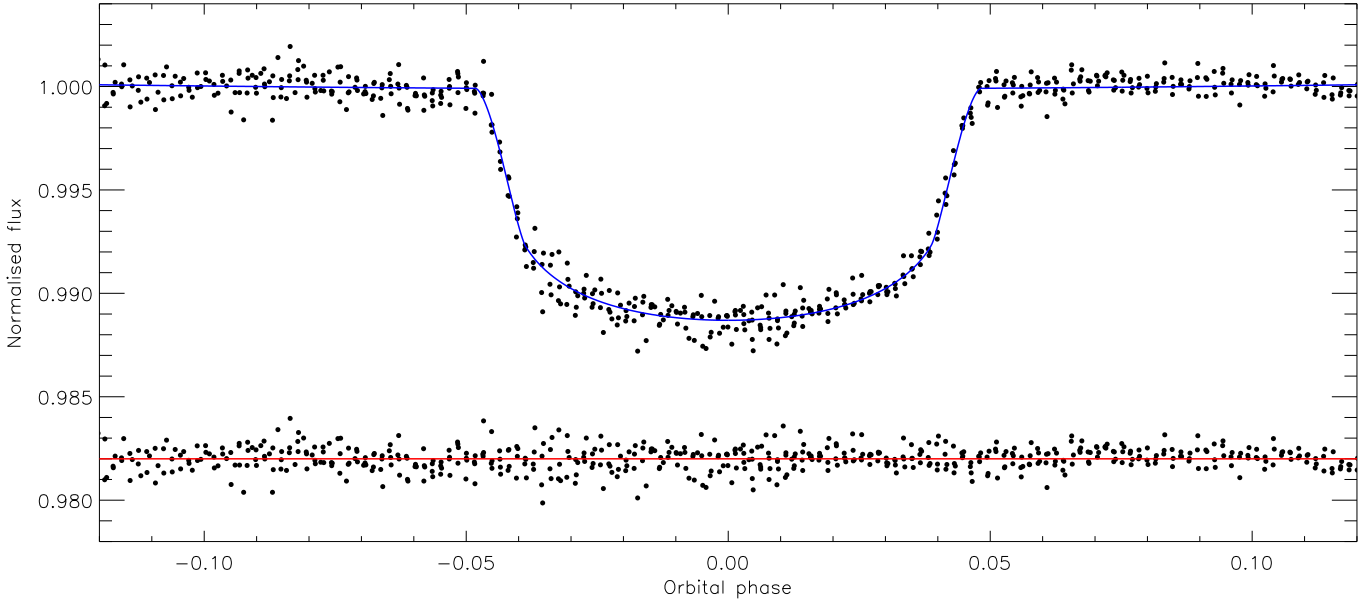}
\caption{\label{fig:lcfit} Phased light curve of WASP-18 and the best fit
found using {\sc jktebop} and the quadratic LD law. The residuals of the fit
are offset from zero to appear at the base of the figure.\\} \end{figure*}

The light curves of WASP-18 were analysed with the {\sc jktebop}%
%-------------------------------------------footnote---------------------------------------------
\footnote{{\sc jktebop} is written in {\sc fortran77} and the source code is available at {\tt http://www.astro.keele.ac.uk/$\sim$jkt/}}
%-------------------------------------------footnote---------------------------------------------
code \citep{Me+04mn,Me+04mn2}, using the approach discussed in detail by \citet{Me08mn}. {\sc jktebop} approximates the two components of WASP-18 using biaxial spheroids whose shapes are governed by the mass ratio; we adopt the value 0.008 but find that large changes in this number have a negligible effect on the results.

WASP-18\,b has a slightly eccentric orbit, which should be taken into account as it has a small effect on the shape of the transit. However, for transit light curves the effect is in general too subtle to include it as a fitted parameter \citep{Kipping08mn}. We could fix eccentricity, $e$, and periastron longitude, $\omega$, to the values obtained from the velocity variation of the parent star (H09), but this would neglect their measurement uncertainties. We have therefore modified {\sc jktebop} to allow the inclusion of $e$ and $\omega$ as fitted parameters constrained by the known values and uncertainties. In practise we use $e \cos \omega = 0.0008 \pm 0.0014$ and $e \sin \omega = -0.0093 \pm 0.0030$, as these two parameters are only weakly correlated with each other. In the current case the uncertainties in the $e \cos \omega$ and $e \sin \omega$ values have a very minor effect on our results.

We have incorporated the H09 photometry of WASP-18 in order to obtain the most precise ephemeris. This was done by including the reference transit epoch from H09 as an observed quantity, and then fitting for transit epoch and orbital period directly \citep[see][]{Me+07aa}. We chose a new transit epoch which is close to the midpoint of our own observations, so is essentially based on just the data presented in this work. The final eclipse ephemeris is
$$ T_0 = {\rm BJD} \,\, 2\,455\,084.792931 (88) \, + \, 0.94145181 (44) \times E $$
where $T_0$ is the transit midpoint, $E$ is the number of orbital cycles after the reference epoch, and quantities in parentheses denote the uncertainty in the final digit of the preceding number.

% \begin{table*} \centering \caption{\label{tab:lcfinal} Final parameters of the
% fit to the light curve of WASP-18, based on the individual solutions with both LD
% coefficients included as fitted parameters (Table\,\ref{tab:lcfit}). The results
% found by \citet{Hellier+09natur} are included for comparison (without errorbars
% if they were not directly quoted quantities.}
% % \setlength{\tabcolsep}{4.5pt}
% \begin{tabular}{l r@{\,$\pm$\,}l r@{\,$\pm$\,}l}
% \hline \hline
% \                     &    \mc{This work}        & \mc{\citet{Hellier+09natur}} \\
% \hline
% Orbital period (d)    & 0.94145182& 0.00000044   & 0.94145299 & 0.00000087      \\
% $T_0$ (HJD)           & 2455084.792931& 0.000084 & 2454221.48163 & 0.00038      \\
% $r_{\rm A}+r_{\rm b}$ &      0.3067  & 0.0096    & \mc{0.3053}                  \\
% $k$                   &      0.0973  & 0.0016    & 0.0935 & 0.0011              \\
% $i$ ($^\circ$)        &     85.0     & 2.1       & 86.0 & 2.5                   \\
% $r_{\rm A}$           &      0.2795  & 0.0084    & \mc{0.2792}                  \\
% $r_{\rm b}$           &      0.0272  & 0.0012    & \mc{0.0261}                  \\
% \hline \hline \end{tabular} \end{table*}

\begin{table*} \caption{\label{tab:lcfits} Parameters of the {\sc jktebop} best fits of
our $V$-band light curve of WASP-18, using different approaches to LD. For each part of
the table the upper quantities are fitted parameters and the lower quantities are derived
parameters. $r_{\rm A}$ and $r_{\rm b}$ are the fractional radii of the star and planet,
respectively, and $k = r_{\rm b}/r_{\rm A}$. $i$ is the orbital inclination, and $u_{\rm A}$
and $v_{\rm A}$ are the linear and non-linear LD coefficients, respectively. $P$ is the
orbital period and $T_0$, the reference epoch of minimum light, is given as BJD $-$ 2455000.0.}
\begin{tabular}{l r@{\,$\pm$\,}l r@{\,$\pm$\,}l r@{\,$\pm$\,}l r@{\,$\pm$\,}l r@{\,$\pm$\,}l}
\hline \hline
\                     &      \mc{Linear LD law}     &    \mc{Quadratic LD law}    &   \mc{Square-root LD law}   &   \mc{Logarithmic LD law}   &     \mc{Cubic LD law}       \\
\hline
\multicolumn{11}{l}{All LD coefficients fixed} \\
\hline
$r_{\rm A}+r_{\rm b}$ & 0.3094       & 0.0075       & 0.3034       & 0.0074       & 0.3060       & 0.0077       & 0.3069       & 0.0075       & 0.3176       & 0.0068       \\
$k$                   & 0.09704      & 0.00075      & 0.09636      & 0.00057      & 0.09682      & 0.00069      & 0.09712      & 0.00062      & 0.09912      & 0.00048      \\
$i$ (deg.)            & 84.68        &  1.33        & 85.96        &  1.70        & 85.31        &  1.51        & 85.05        &  1.35        & 83.05        &  0.90        \\
$u_{\rm A}$           &      \mc{ 0.60 fixed}       &      \mc{ 0.40 fixed}       &      \mc{ 0.20 fixed}       &      \mc{ 0.70 fixed}       &      \mc{ 0.40 fixed}       \\
$v_{\rm A}$           &            \mc{ }           &      \mc{ 0.30 fixed}       &      \mc{ 0.60 fixed}       &      \mc{ 0.25 fixed}       &      \mc{ 0.15 fixed}       \\
$P$                   & 0.94145182   & 0.00000040   & 0.94145181   & 0.00000041   & 0.94145182   & 0.00000044   & 0.94145181   & 0.00000045   & 0.94145181   & 0.00000043   \\
$T_0$                 & 84.792935    &  0.000090    & 84.792929    &  0.000085    & 84.792932    &  0.000084    & 84.792931    &  0.000084    & 84.792924    &  0.000085    \\
\hline
$r_{\rm A}$           & 0.2821       & 0.0067       & 0.2767       & 0.0066       & 0.2790       & 0.0069       & 0.2798       & 0.0067       & 0.2890       & 0.0060       \\
$r_{\rm b}$           & 0.02737      & 0.00085      & 0.02667      & 0.00078      & 0.02701      & 0.00085      & 0.02717      & 0.00081      & 0.02864      & 0.00072      \\
$\sigma$ ($m$mag)     &        \mc{ 0.6406}         &        \mc{ 0.6358}         &        \mc{ 0.6356}         &        \mc{ 0.6355}         &        \mc{ 0.6408}         \\
$\chi^2_{\ \nu}$      &        \mc{ 1.0194}         &        \mc{ 1.0048}         &        \mc{ 1.0031}         &        \mc{ 1.0034}         &        \mc{ 1.0247}         \\
\hline
\multicolumn{11}{l}{Fitting for the linear LD coefficient and fixing the nonlinear LD coefficient} \\
\hline
$r_{\rm A}+r_{\rm b}$ & 0.3106       & 0.0070       & 0.3053       & 0.0082       & 0.3073       & 0.0080       & 0.3066       & 0.0076       & 0.3071       & 0.0079       \\
$k$                   & 0.09813      & 0.00070      & 0.09672      & 0.00087      & 0.09720      & 0.00082      & 0.09702      & 0.00081      & 0.09726      & 0.00086      \\
$i$ (deg.)            & 84.2         &  1.1         & 85.5         &  1.8         & 85.0         &  1.5         & 85.1         &  1.5         & 85.0         &  1.5         \\
$u_{\rm A}$           & 0.527        & 0.023        & 0.384        & 0.029        & 0.180        & 0.024        & 0.705        & 0.025        & 0.495        & 0.025        \\
$v_{\rm A}$           &            \mc{ }           &      \mc{ 0.30 fixed}       &      \mc{ 0.60 fixed}       &      \mc{ 0.25 fixed}       &      \mc{ 0.15 fixed}       \\
$P$                   & 0.94145182   & 0.00000043   & 0.94145181   & 0.00000041   & 0.94145182   & 0.00000042   & 0.94145181   & 0.00000042   & 0.94145182   & 0.00000039   \\
$T_0$                 & 84.792931    &  0.000080    & 84.792930    &  0.000089    & 84.792931    &  0.000086    & 84.792931    &  0.000081    & 84.792932    &  0.000084    \\
\hline
$r_{\rm A}$           & 0.2829       & 0.0062       & 0.2784       & 0.0072       & 0.2801       & 0.0071       & 0.2795       & 0.0068       & 0.2799       & 0.0070       \\
$r_{\rm b}$           & 0.02776      & 0.00078      & 0.02692      & 0.00094      & 0.02722      & 0.00090      & 0.02711      & 0.00086      & 0.02722      & 0.00088      \\
$\sigma$ ($m$mag)     &        \mc{ 0.6357}         &        \mc{ 0.6356}         &        \mc{ 0.6352}         &        \mc{ 0.6355}         &        \mc{ 0.6350}         \\
$\chi^2_{\ \nu}$      &        \mc{ 1.0048}         &        \mc{ 1.0061}         &        \mc{ 1.0037}         &        \mc{ 1.0049}         &        \mc{ 1.0031}         \\
\hline
\multicolumn{11}{l}{Fitting for the linear LD coefficient and perturbing the nonlinear LD coefficient} \\
\hline
$r_{\rm A}+r_{\rm b}$ & 0.3106       & 0.0066       & 0.3053       & 0.0084       & 0.3073       & 0.0079       & 0.3066       & 0.0081       & 0.3071       & 0.0079       \\
$k$                   & 0.09813      & 0.00069      & 0.09672      & 0.00092      & 0.09720      & 0.00081      & 0.09702      & 0.00086      & 0.09726      & 0.00083      \\
$i$ (deg.)            & 84.2         &  1.0         & 85.5         &  1.8         & 85.0         &  1.5         & 85.1         &  1.6         & 85.0         &  1.5         \\
$u_{\rm A}$           & 0.527        & 0.022        & 0.384        & 0.039        & 0.180        & 0.042        & 0.705        & 0.053        & 0.495        & 0.028        \\
$v_{\rm A}$           &            \mc{ }           &     \mc{ 0.30 perturbed}    &     \mc{ 0.60 perturbed}    &     \mc{ 0.25 perturbed}    &     \mc{ 0.15 perturbed}    \\
$P$                   & 0.94145182   & 0.00000042   & 0.94145181   & 0.00000043   & 0.94145182   & 0.00000043   & 0.94145181   & 0.00000043   & 0.94145182   & 0.00000043   \\
$T_0$                 & 84.792931    &  0.000086    & 84.792930    &  0.000085    & 84.792931    &  0.000080    & 84.792931    &  0.000084    & 84.792932    &  0.000084    \\
\hline
$r_{\rm A}$           & 0.2829       & 0.0059       & 0.2784       & 0.0075       & 0.2801       & 0.0070       & 0.2795       & 0.0072       & 0.2799       & 0.0070       \\
$r_{\rm b}$           & 0.02776      & 0.00076      & 0.02692      & 0.00095      & 0.02722      & 0.00086      & 0.02711      & 0.00090      & 0.02722      & 0.00091      \\
$\sigma$ ($m$mag)     &        \mc{ 0.6357}         &        \mc{ 0.6356}         &        \mc{ 0.6352}         &        \mc{ 0.6355}         &        \mc{ 0.6350}         \\
$\chi^2_{\ \nu}$      &        \mc{ 1.0048}         &        \mc{ 1.0061}         &        \mc{ 1.0037}         &        \mc{ 1.0049}         &        \mc{ 1.0031}         \\
\hline
\multicolumn{11}{l}{Fitting for both LD coefficients} \\
\hline
$r_{\rm A}+r_{\rm b}$ & 0.3106       & 0.0070       & 0.3064       & 0.0086       & 0.3068       & 0.0083       & 0.3063       & 0.0083       & 0.3071       & 0.0081       \\
$k$                   & 0.09813      & 0.00070      & 0.09740      & 0.00117      & 0.09721      & 0.00144      & 0.09737      & 0.00128      & 0.09721      & 0.00140      \\
$i$ (deg.)            & 84.2         &  1.1         & 85.0         &  1.7         & 85.0         &  1.9         & 85.0         &  1.7         & 85.0         &  1.8         \\
$u_{\rm A}$           & 0.527        & 0.022        & 0.473        & 0.087        & 0.213        & 0.433        & 0.617        & 0.175        & 0.492        & 0.045        \\
$v_{\rm A}$           &            \mc{ }           & 0.11         & 0.18         & 0.54         & 0.77         & 0.12         & 0.24         & 0.16         & 0.20         \\
$P$                   & 0.94145182   & 0.00000041   & 0.94145181   & 0.00000044   & 0.94145181   & 0.00000043   & 0.94145182   & 0.00000044   & 0.94145181   & 0.00000041   \\
$T_0$                 & 84.792931    &  0.000086    & 84.792932    &  0.000089    & 84.792931    &  0.000084    & 84.792932    &  0.000088    & 84.792931    &  0.000084    \\
\hline
$r_{\rm A}$           & 0.2829       & 0.0062       & 0.2792       & 0.0076       & 0.2796       & 0.0072       & 0.2792       & 0.0073       & 0.2799       & 0.0071       \\
$r_{\rm b}$           & 0.02776      & 0.00078      & 0.02719      & 0.00103      & 0.02718      & 0.00108      & 0.02718      & 0.00103      & 0.02721      & 0.00105      \\
$\sigma$ ($m$mag)     &        \mc{ 0.6357}         &        \mc{ 0.6353}         &        \mc{ 0.6352}         &        \mc{ 0.6354}         &        \mc{ 0.6350}         \\
$\chi^2_{\ \nu}$      &        \mc{ 1.0048}         &        \mc{ 1.0055}         &        \mc{ 1.0053}         &        \mc{ 1.0057}         &        \mc{ 1.0047}         \\
\hline \hline\\ \end{tabular} \end{table*}

The limb darkening (LD) of WASP\,18\,A was accounted for using five different parametric laws \citep[see][]{Me08mn}. Theoretical LD coefficients were obtained by bilinear interpolation, to the known effective temperature (\Teff) and surface gravity of the parent star, in the tables of \citet{Vanhamme93aj}, \citet{Claret00aa,Claret04aa2} and \citet{ClaretHauschildt03aa}. We obtained solutions with the LD coefficients fixed at the theoretical values, with the linear coefficient fitted for and the nonlinear coefficient fixed (and optionally perturbed by $\pm$0.05 on a flat distribution in the error analyses), and with both LD coefficients included as fitted parameters. The full set of solutions is given in Table\,\ref{tab:lcfits}.

The uncertainties of the light curve parameters were assessed using Monte Carlo simulations \citep{Me+04mn3,Me+05mn}. The importance of red noise was checked using a residual-permutation approach, and found to be minor. The solutions with LD coefficients fixed to theoretical values are poorer than those where one or both LD coefficients are fitted parameters. \reff{We adopt the mean of the solutions with non-linear LD and both LD coefficients fitted, as these are the most internally consistent}. The final uncertainties come from Monte Carlo solutions but include contributions from the residual-permutation analyses and the (minor) variation between the solutions with different LD laws. We find the fractional radius\footnote{Fractional radius is the radius of a component of a binary system expressed as a fraction of the orbital semimajor axis. The utility of this quantity is that it is measureable from light curve data alone.} of the star and planet to be $r_{\rm A} = 0.2795 \pm 0.0084$ and $r_{\rm b} = 0.0272 \pm 0.0012$, respectively, and the orbital inclination to be $i = 85.0^\circ \pm 2.1^\circ$.

% \begin{table*} \centering \caption{\label{tab:lcfinal} Final parameters of the
% fit to the light curve of WASP-18, based on the individual solutions with both LD
% coefficients included as fitted parameters (Table\,\ref{tab:lcfit}). The results
% found by \citet{Hellier+09natur} are included for comparison (without errorbars
% if they were not directly quoted quantities.}
% % \setlength{\tabcolsep}{4.5pt}
% \begin{tabular}{l r@{\,$\pm$\,}l r@{\,$\pm$\,}l}
% \hline \hline
% \                     &    \mc{This work}        & \mc{\citet{Hellier+09natur}} \\
% \hline
% Orbital period (d)    & 0.94145182& 0.00000044   & 0.94145299 & 0.00000087      \\
% $T_0$ (HJD)           & 2455084.792944& 0.000084 & 2454221.48163 & 0.00038      \\
% $r_{\rm A}+r_{\rm b}$ &      0.3067  & 0.0096    & \mc{0.3053}                  \\
% $k$                   &      0.0973  & 0.0016    & 0.0935 & 0.0011              \\
% $i$ ($^\circ$)        &     85.0     & 2.1       & 86.0 & 2.5                   \\
% $r_{\rm A}$           &      0.2795  & 0.0084    & \mc{0.2792}                  \\
% $r_{\rm b}$           &      0.0272  & 0.0012    & \mc{0.0261}                  \\
% \hline \hline \end{tabular} \end{table*}

%%%%%%%%%%%%%%%%%%%%%%%%%%%%%%%%%%%%%%%%%%%%%%%%%%%%%%%%%%%%%%%%%%%%%%%%%%%%%%%%%%%%%%%%%%%%%%%%%%%

\section{The physical properties of WASP-18}

\begin{table*} \centering
\caption{\label{tab:absdimall} Physical properties for WASP-18, derived
using the predictions of different sets of stellar evolutionary models.
% Mass, radius, surface gravity and density are denoted by $M$, $R$, $g$ and $\rho$,
% respectively, and are subscripted with an ${\rm A}$ to indicate the star or with
% a ${\rm b}$ to indicate the planet.
For quantities with two errorbars, the
uncertainties have been split into statistical and systematic errors, respectively.
}
\begin{tabular}{l l r@{\,$\pm$\,}l r@{\,$\pm$\,}l r@{\,$\pm$\,}l r@{\,$\pm$\,}c@{\,$\pm$\,}l r@{\,$\pm$\,}l}
\hline \hline
\ & & \mc{{\it Cambridge} models}  & \mc{{\it Y$^2$} models} & \mc{{\it Claret} models}
            & \multicolumn{3}{c}{\bf Final result (this work)} & \mc{\citet{Hellier+09natur}} \\
\hline
Orbital separation& (AU)    &0.02022&0.00022&0.02043&0.00028&0.02047&0.00027& {\bf 0.02047}& {\bf 0.00028}& {\bf 0.00025}& 0.02026 & 0.00068         \\[2pt]
Stellar mass      & (\Msun) & 1.235 & 0.039 & 1.274 & 0.052 & 1.281 & 0.050 & {\bf 1.281}  & {\bf 0.052}  & {\bf 0.046}  & 1.25 & 0.13               \\
Stellar radius    & (\Rsun) & 1.215 & 0.048 & 1.227 & 0.043 & 1.230 & 0.042 & {\bf 1.230}  & {\bf 0.045}  & {\bf 0.015}  & \erm{1.216}{0.067}{0.054} \\
Stellar \logg     & [cgs]   & 4.361 & 0.022 & 4.365 & 0.026 & 4.366 & 0.026 & {\bf 4.366}  & {\bf 0.026}  & {\bf 0.005}  & \erm{4.367}{0.028}{0.042} \\[2pt]
Stellar density   & (\psun) & 0.689 & 0.062 & 0.689 & 0.062 & 0.689 & 0.062 & {\bf 0.689}  & {\bf 0.062}  & {\bf 0.000}  & \erm{0.707}{0.056}{0.096} \\[2pt]
Planetary mass    & (\Mjup) & 10.18 & 0.22  & 10.39 & 0.30  & 10.43 & 0.28  & {\bf 10.43}  & {\bf 0.30 }  & {\bf 0.24 }  & 10.30 & 0.69              \\
Planetary radius  & (\Rjup) & 1.151 & 0.052 & 1.163 & 0.055 & 1.165 & 0.054 & {\bf 1.165}  & {\bf 0.055}  & {\bf 0.014}  & \erm{1.106}{0.072}{0.054} \\
Planet surface gravity & (\mss)& 191 & 17   &  191  &  17   &  191  &  17   & {\bf 191}    & \multicolumn{2}{l}{\bf 17}  & \erm{194}{12}{21}         \\
Planetary density & (\pjup) & 6.68  & 0.89  & 6.61  & 0.89  &  6.60 & 0.88  & {\bf 6.60}   & {\bf 0.90}   & {\bf 0.08}   & \erm{7.73}{0.78}{1.27}    \\[2pt]
Stellar age       & (Gyr)   &\mc{0.0 -- 0.6}&\mc{0.0 -- 2.1}&\erm{0.4}{1.2}{0.4} & \multicolumn{3}{c}{\bf 0.0 -- 2.0}    & \mc{0.5 -- 1.5}           \\
\hline \hline\\ \end{tabular} \end{table*}

% \begin{table*}
% \caption{\label{tab:absdim} Final physical properties for WASP-18. The first
% error bars are statistical and the second are systematic. The corresponding
% results from \citet{Hellier+09natur} have been included for comparison.}
% \begin{tabular}{l l r@{\,$\pm$\,}c@{\,$\pm$\,}l r@{\,$\pm$\,}l} \hline \hline
%     &   & \multicolumn{3}{c}{Final result (this work)} & \citet{Hellier+09natur} \\
% \hline
% $a$ & (AU)               & 0.02047& 0.00028& 0.00039 & 0.2026 & 0.00068          \\
% $M_{\rm A}$ & (\Msun)    & 1.281  & 0.052  & 0.072   & 1.25 & 0.13               \\
% $R_{\rm A}$ & (\Rsun)    & 1.023  & 0.045  & 0.072   & \erm{1.216}{0.067}{0.054} \\
% $\log g_{\rm A}$ & [cgs] & 4.366  & 0.026  & 0.008   & \erm{4.367}{0.028}{0.042} \\
% $\rho_{\rm A}$ & (\psun) & 0.689  & 0.062  & 0.000   & \erm{0.707}{0.056}{0.096} \\
% $M_{\rm b}$ & (\Mjup)    & 10.43  & 0.30   & 0.39    & 10.30 & 0.69              \\
% $R_{\rm b}$ & (\Rjup)    & 1.165  & 0.055  & 0.023   & \erm{1.106}{0.072}{0.054} \\
% $g_{\rm b}$ & (\mss)     & 191 &\multicolumn{2}{l}{17}&\erm{194}{12}{21}         \\
% $\rho_{\rm b}$ & (\pjup) & 6.60   & 0.90   & 0.12    & \erm{7.73}{0.78}{1.27}    \\
% \hline \hline \end{tabular} \end{table*}

The physical properties of a transiting planetary system cannot in general be calculated purely from observed quantities. The most common way to overcome this difficulty is to impose predictions from theoretical stellar evolutionary models onto the parent star. We have used tabulated predictions from three sources: {\it Claret} \citep{Claret04aa, Claret05aa, Claret06aa2, Claret07aa2}, {\it Y$^2$} \citep{Demarque+04apjs} and {\it Cambridge} \citep{Pols+98mn,EldridgeTout04mn}. This allows the assessment of the systematic errors caused by using stellar theory.

We began with the parameters measured from the light curve and the observed velocity amplitude of the parent star, $K_{\rm A} = 1818.3 \pm 8.0$\,m\,s$^{-1}$ (H09). These were augmented by an estimate of the velocity amplitude of the {\em planet}, $K_{\rm b}$, to calculate preliminary physical properties of the system. We then interpolated within one of the grids of theoretical predictions to find the expected radius and \Teff\ of the star for the preliminary mass and the measured metal abundance ($\FeH = 0.00 \pm 0.09$; H09). $K_{\rm b}$ was then iteratively refined to minimise the difference between the model-predicted radius and \Teff, and the calculated radius and measured \Teff\ ($6400 \pm 100$\,K; H09). This was done for a range of ages for the star, and the best overall fit retained as the optimal solution. Finally, the above process was repeated whilst varying every input parameter by its uncertainty to build up a complete error budget for each output parameter \citep{Me+05aa}. A detailed description of this process can be found in \citet{Me09mn}.

Table\,\ref{tab:absdimall} shows the results of these analyses. The physical properties calculated using the {\it Claret} and {\it Y$^2$} sets of stellar models are in excellent agreement, but those from the {\it Cambridge} models are slightly discrepant. This causes systematic errors of 4\% in the stellar mass and 2\% in the planetary mass, both of similar size to the corresponding statistical errors. The quality of our results is therefore limited by our theoretical understanding of the parent star. Our final results are in good agreement with those of H09 (Table\,\ref{tab:absdimall}), but incorporate a more comprehensive set of uncertainties.

These results give the equilibrium temperature of the planet to be one of the highest for the known planets:
$$ T_{\rm eq} \ = \ (2392 \pm 63) \ \left(\frac{1-A}{4F}\right)^{1/4} \ \ {\rm K} $$
where $A$ is the Bond albedo and $F$ is the heat redistribution factor. This equilibrium temperature, and the closeness to its parent star, make WASP-18\,b a good target for the detection of thermal emission and reflected light.

%%%%%%%%%%%%%%%%%%%%%%%%%%%%%%%%%%%%%%%%%%%%%%%%%%%%%%%%%%%%%%%%%%%%%%%%%%%%%%%%%%%%%%%%%%%%%%%%%%%

\section{Conclusions}

We have presented high-quality observations of five consecutive transits by the newly-discovered planet WASP-18\,b, which has \reff{one of the shortest orbital periods} of all known transiting extrasolar planetary systems (TEPs). Our defocussed-photometry approach yielded scatters of between 0.47 and 0.83 mmag per point in the final light curves. These data were analysed using the {\sc jktebop} code, which was modified to include the spectroscopically derived orbital eccentricity in a statistically correct way. The light curve parameters were then combined with the predictions of theoretical stellar evolutionary models to determine the physical properties of the planet and its host star.

A significant source of uncertainty in our results stems from the use of theoretical models to constrain the physical properties of the star. Further uncertainty comes from observed \Teff\ and \MoH, for which improved values are warranted. However, the systematic error from the use of stellar theory is an important uncertainty in the masses of the star and planet. This is due to our fundamentally incomplete understanding of the structure and evolution of low-mass stars. As with many other transiting systems (e.g.\ WASP-4; \citealt{Me+09mn2}), our understanding of the planet is limited by our lack of understanding of the parent star.

\reff{We confirm and refine the physical properties of WASP\,18 found by H09.} WASP-18\,b is a very massive planet in an extremely short-period and {\em eccentric} orbit, which is a clear indicator that the tidal effects in planetary systems are weaker than expected (see H09). Long-term follow-up studies of WASP-18 will add progressively stricter constraints on the orbital decay of the planet and thus the strength of these tidal effects.

We now split the full sample of known (i.e.\ published) TEPS into two classes according to planetary mass. The mass distribution of transiting planets shows a dearth of objects with masses in the interval 2.0--3.1\Mjup. There are nine planets more massive than this and 46 less massive. Seven of the nine high-mass TEPs have eccentric orbits (HAT-P-2, \citealt{Bakos+07apj3}; HD\,17156, \citealt{Barbieri+07aa}; HD\,80606, \citealt{Laughlin+09natur}; WASP-10, \citealt{Christian+09mn}; WASP-14, \citealt{Joshi+09mn}; WASP-18; XO-3, \citealt{Johnskrull+08apj}), and the existing radial velocity observations of the remaining two cannot rule out an eccentricity of $0.03$ or lower (CoRoT-Exo-2, \citealt{Alonso+08aa2}; OGLE-TR-L9, \citealt{Snellen+09aa}). By comparison, only four of the 46 low-mass TEPs have a {\em significant} \citep{LucySweeney71aj} orbital eccentricity measurement.

These numbers imply that the more massive TEPs are a different population to the less massive ones; Fisher's exact test \citep{Fisher22jrss} returns a probability lower than $10^{-5}$ of the null hypothesis (although this does not account for our freedom to choose the dividing line between the two classes). This indicates that the two types of TEPs have a different \reff{internal structure, formation mechanism, or evolution,} a suggestion which is supported by observations of misalignment between the spin and orbital axes of $M > 3$\Mjup\ TEPs \citep{Johnson+09pasp}.

There is, however, a bias at work here. The more massive TEPs cause a larger radial velocity signal in their parent star ($M_{\rm b} \propto K^3$), so a given set of radial velocity measurements can detect smaller eccentricities \citep[see also][]{ShenTurner08apj}. The eccentricity of the WASP-18 system is in fact below the detection limit of existing observations of most TEPs. We therefore advocate the acquisition of additional velocity data for the known low-mass TEPs, in order to equalise the eccentricity detection limits between the two classes of TEPs. These observations would allow acceptance or rejection of the hypothesis that more massive TEPs represent a fundamentally different planet population to their lower-mass brethren.

% WASP-18b: M=10.30 Mjup,R=1.106Rjup, rho=7.74rhojup, Teq=2384. Interesting.
% Other massive TEPs:     e
% CoRoT-Exo-2  3.31       0.0             constraint: e=0.03pm0.03
% HAT-P-2      8.72       0.5170
% HD-17156     3.212      0.6753
% HD-80606     4.20       0.9329
% OGLE-TR-L9   4.5        0.0             almost unconstrained
% WASP-10      3.15       0.057
% WASP-14      7.34       0.091
% WASP-18      10.4       0.0092
% XO-3         11.79      0.2884

%%%%%%%%%%%%%%%%%%%%%%%%%%%%%%%%%%%%%%%%%%%%%%%%%%%%%%%%%%%%%%%%%%%%%%%%%%%%%%%%%%%%%%%%%%%%%%%%%%%

\acknowledgments

The observations presented in this work will be made available at the CDS ({\tt http://cdsweb.u-strasbg.fr/}) and at {\tt http://www.astro.keele.ac.uk/$\sim$jkt/}. The operation of the Danish 1.54m telescope was financed by the Danish Natural Science Research Council (FNU). We thank Dr.\ J.\ Eldridge for calculating the {\it Cambridge} set of stellar models used in this work. J\,Southworth and DRA acknowledge financial support from STFC in the form of postdoctoral research assistant positions. Astronomical research at the Armagh Observatory is funded by the Northern Ireland Department of Culture, Arts and Leisure (DCAL). DR (boursier FRIA), FF and J\,Surdej acknowledge support from the Communaut\'e fran\c{c}aise de Belgique -- Actions de recherche concert\'ees -- Acad\'emie Wallonie-Europe. The following internet-based resources were used in research for this paper: the ESO Digitized Sky Survey; the NASA Astrophysics Data System; the SIMBAD database operated at CDS, Strasbourg, France; and the ar$\chi$iv scientific paper preprint service operated by Cornell University.

% \bibliographystyle{mn_new}
% \bibliography{aamnem99,jkt}
% \bsp

\end{document}